\documentclass[a4paper,11pt]{article}
\usepackage{jheppub} 
\usepackage{lineno}
\usepackage{multirow}
\pdfoutput=1
\usepackage{booktabs}
\usepackage{subcaption}
\usepackage{tabularx,array,booktabs,multirow,bigstrut}
\usepackage{siunitx}
\usepackage{float}
\usepackage[compat=1.1.0]{tikz-feynman}
\usepackage{indentfirst}
\usepackage{amssymb}
\usepackage{comment}
\usepackage[utf8]{inputenc}
\usepackage[T1]{fontenc}

\makeatletter

\newcommand{\Rmnum}[1]{\expandafter\@slowromancap\romannumeral #1@}
\makeatother

\arxivnumber{2405.09417} 

\title{\boldmath  Study of charged Lepton Flavor Violation in electron muon interactions}

\author[a]{Ran Ding,}
\author[a]{Jingshu Li,}
\author[a]{Meng Lu,}
\author[a]{Zhengyun You,}
\author[b]{Zijian Wang,}
\author[b]{and Qiang Li}

\affiliation[a]{
School of Physics, Sun Yat-Sen University, Guangzhou 510275, China}
\affiliation[b]{
School of Physics and State Key Laboratory of Nuclear Physics and Technology, Peking University, Beijing, 100871, China}

\emailAdd{dingr25@mail2.sysu.edu.cn}
\emailAdd{lijsh53@mail2.sysu.edu.cn}
\emailAdd{lumeng5@mail.sysu.edu.cn}
\emailAdd{youzhy5@mail.sysu.edu.cn}
\emailAdd{wangzijian@stu.pku.edu.cn}
\emailAdd{qliphy0@pku.edu.cn}

\abstract{With the improvement of muon acceleration technology, it has received great interests to exploit high-energy muon beams for collision or target experiments. We investigate the possible charged Lepton Flavor Violation~(cLFV) processes mediated by an extra massive neutral gauge boson $Z'$ in electron muon interactions, either at a proposed electron muon collider or in a fixed target experiment with high-energy muon beam hitting electrons in the target. Based on Monte Carlo calculations and fast detector simulations, we study in detail the signal and possible backgrounds, giving the sensitivity results of cLFV signals at the 90\% confidence level. Compared with the current and prospective limits set by other experiments, electron muon interactions demonstrate significant advantages in the cLFV coupling strength sensitivity with $\tau$ in the final states. In addition, a special cLFV coupling combination, $\lambda _{e\mu} \times \lambda _{\mu\mu}$, can also be probed in our proposal.}


\begin{document}
\maketitle
\flushbottom

\section{Introduction}
\label{sec:intro}

Collider experiment serves as a crucial tool for precision measurement of the standard model~(SM) and search for new physics beyond the SM (BSM), with the experimental technology constantly being improved and evolved. In the near future, the High-Luminosity Large Hadron Collider~(HL-LHC)~\cite{Apollinari:2017lan}, the Future Circular Collider~(FCC)~\cite{FCC:2018byv,FCC:2018evy,FCC:2018vvp} or the Circular Electron Positron Collider~(CEPC)~\cite{CEPCStudyGroup:2018rmc,CEPCStudyGroup:2018ghi} may become important instruments for the next generation high energy frontier research.
While recently with the continuous development of muon acceleration technology, the muon collider has also become an increasingly popular consideration. Since it integrates the advantages of electron colliders and hadron colliders, the muon collider may become a golden factory for studying various new physics processes~\cite{Aime:2022flm}. 

On the other hand, high energy muon beams can also be used to create electron muon collisions. As early as ten to thirty years ago, numerous research efforts have already focused on the potential of electron muon collider~\cite{Barger:1997dv,Choi:1997bm,Montero:1998sv,Cvetic:1999fk,Almeida:2000qd,Singhal:2007hw}. In recent years, as the construction of high energy lepton collider gradually became feasible from the engineering standpoint, interest in electron-muon collision has been reignited~\cite{Bouzas:2023vba,Bossi:2020yne,Lu:2020dkx,Bouzas:2021sif}. 

Recently, a new collider proposal, $\mu$TRISTAN, has been proposed based on an ultra-cold muon technology developed for the muon~$(g-2)$ experiment at J-PARC~\cite{Hamada:2022mua}. It includes a $\mu ^+ \mu ^+$ collider and a $\mu ^+ e^-$ collider, in which we are interested in the later. The main parameters from the $\mu$TRISTAN $e\mu$ collider proposal~\cite{Hamada:2022mua,Akturk:2024jbl} are listed in Tab.~\ref{tab:muTRISTAN}. According to several phenomenological study, $\mu$TRISTAN may have certain potentials on measurements related to Higgs and new physics searches~\cite{Das:2024gfg,Lichtenstein:2023iut,Hamada:2022uyn}.

\begin{table}[!t]
\centering
\caption{The main parameters of the $\mu$TRISTAN $e\mu$ collider.}
\label{tab:muTRISTAN}
\begin{tabular}{c c c c }
\toprule
\midrule
Parameter&Electron&Anti-muon\\
\midrule
Beam energy&30~GeV&1~TeV\\
Polarization&70\%&$>$25\%\\
Particles per bunch~($10^{10}$)&6.2&1.4\\
\midrule
Luminosity&\multicolumn{2}{c}{4.6×10$^{33}$~cm$^{-1}$s$^{-1}$}\\
Collision frequency&\multicolumn{2}{c}{4×10$^6$~Hz}\\
\midrule
\bottomrule
\end{tabular}
\end{table}

Meanwhile, utilizing high-density muon beams to strike fixed targets can also provide a possibility to search for new physics. Many such attempts have been proposed, including the Muon Missing Momentum~($M^{3}$) proposal at Fermilab~\cite{kahn2018m3}, and a recent idea of searching for muonic force carriers by using ATLAS detector as a fixed target~\cite{galon2020searching}. 

In this study we also investigate a fixed electron-target experiment with a muon beam in addition to the $e\mu$ collider. As lepton flavor in the initial state is non-zero, electron muon interaction can strongly avoid many potential background processes which would occur at different-sign muon colliders or electron-positron colliders, thus possessing higher sensitivity to new physics signals, typically the charged Lepton Flavor Violation (cLFV) processes.

In the SM framework, the cLFV processes are strongly suppressed due to the tiny mass of neutrinos, hence unobservable in the current experiments yet. However, it may be much enhanced in various BSM models, such as super-symmetry~(SUSY)~\cite{Barbier:2004ez}, leptoquark~\cite{Dorsner:2016wpm}, two-Higgs-doublet~\cite{Branco:2011iw}, R-parity-violating~(RPV) Minimal Super-symmetric Standard Model~(MSSM)~\cite{Choudhury:1996ia,Chemtob:2004xr,Cai:2024fkl}, and the heavy neutral gauge boson $Z'$~\cite{Langacker:2008yv} studied in this paper.
In the past decades, searches for the cLFV process were performed in different channels with several approaches, typically the high intensity muon-based experiments including $\mu ^+ \to e ^+ \gamma$~(MEG)~\cite{MEG:2016leq}, $\mu ^+ \to e^+ e^+ e^-$~(SINDRUM)~\cite{SINDRUM:1987nra} and $\mu ^- N \to e^- N$~(SINDRUM~\Rmnum{2})~\cite{SINDRUMII:2006dvw,SINDRUMII:1993gxf,SINDRUMII:1996fti,SINDRUMII:1998mwd}, as well as the collider-based searches for cLFV decays of $Z$~\cite{ATLAS:2014vur,OPAL:1995grn,DELPHI:1996iox}, Higgs~\cite{ATLAS:2019pmk,CMS:2017con} and several hadron resonances~\cite{KTeV:2007cvy,CLEO:2008lxu,BaBar:2021loj,BESIII:2022exh}. Meanwhile, there will be continuous new experiments conducted in the near future to constantly improve the existing limits, such as MEG\Rmnum{2}~\cite{MEGII:2018kmf}, Mu3e~\cite{Mu3e:2020gyw}, COMET~\cite{COMET:2018auw} and Mu2e~\cite{Mu2e:2014fns}.


In this study, we consider the cLFV processes based on the interactions of electron and muon in two scenarios: asymmetric electron muon collision at the $e\mu$ collider and fixed electron-target experiment striking with the muon beam. For the former case, the center-of-mass energy includes the energy point of $\mu$TRISTAN and even higher. While for the latter case, we investigate the muon energy around several tens of GeV to test the $Z'$ couplings at the low energy bound.

\section{Physics processes and Monte Carlo simulation}
\label{sec:phy_mc}

\subsection{cLFV in $Z'$ model}
\label{sec:zpmodel}

By introducing an additional $U(1)$ gauge symmetry into the SM framework, it will correspond to a neutral gauge boson $Z'$. Since the heavy neutral gauge bosons are predicted in many BSM models, it may be one of the most motivated extensions of the SM~\cite{Langacker:2008yv,Langacker:2000ju,Buras:2021btx}. 

The full neutral current can be written as
\begin{equation}
    - \mathcal{L}_{\text{NC}} = eJ^{\mu}_{\text{em}}A_{\mu} + g_1J^{(1)\mu}Z^0_{(1),\mu} + g_2J^{(2)\mu}Z^0_{(2),\mu},
\end{equation}
where $Z^0_{(1)}$ and $Z^0_{(2)}$ respectively correspond to SM $SU(2) \times U(1)$ and extra $U(1)$, and both have the same coupling structure. In general, $Z^0_{(1)}$ and $Z^0_{(2)}$ will mix induced by the electroweak breaking, then related to the mass eigenstates $Z$ and $Z'$ through an orthogonal transformation, while it is generally believed that the magnitude of this mixing angle is quite small, on the order of less than several $10^{-3}$~\cite{Langacker:2000ju}.

In this study, we consider that there are not $Z-Z'$ mixing, and the extra $Z'$ current has the same gauge coupling and chiral strength as the SM $Z$, but allows for coupling with leptons of different generations, similarly as Ref.~\cite{ATLAS:2018mrn,Li:2023lin}. The coupling strength of the $Z'$ and leptons can be described by a matrix $\lambda$ as Eq.~\ref{eq:lambda}
\begin{equation}
    \lambda =\left(\begin{array}{lll}
\lambda_{e e} & \lambda_{e \mu} & \lambda_{e \tau} \\
\lambda_{\mu e} & \lambda_{\mu \mu} & \lambda_{\mu \tau} \\
\lambda_{\tau e} & \lambda_{\tau \mu} & \lambda_{\tau \tau}
\end{array}\right).
        \label{eq:lambda}
\end{equation}

\begin{figure}[htbp]
\centering
\subfloat[$\mu - e$ conversion]
{\includegraphics[width=.3\textwidth]{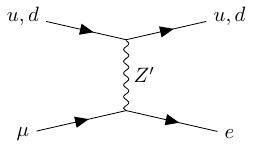}}
\subfloat[$\mu ^+ \to e^+ e^+ e^-$]
{\includegraphics[width=.35\textwidth]{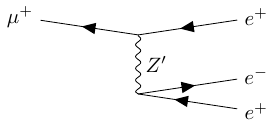}}
\subfloat[$\mu ^+ \to e^+ \gamma$]
{\label{fig:FeynmanothercLFVc}\includegraphics[width=.35\textwidth]{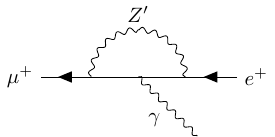}}
\caption{The Feynman diagrams of several $Z'$ mediated cLFV processes.}
\label{fig:FeynmanothercLFV}
\end{figure}

Generally, it represents the strength of the cLFV couplings relative to the SM couplings, assuming that the diagonal elements are 1, while the off-diagonal elements are usually a higher order of magnitude. Therefore, the cases that break lepton flavor conservation twice would not be considered in the further study.
After introducing the $Z'$ boson, the cLFV processes mentioned in Sec.~\ref{sec:intro} can be enhanced by the diagrams shown in Fig.~\ref{fig:FeynmanothercLFV}. The branching ratio limits would be transformed to the coupling $\lambda _{ij}$~\cite{Langacker:2000ju} and compared with our results based on $e\mu$ interaction. 

\subsection{Physics processes}
\label{sec:process}

\begin{figure}[htbp]
\centering
\subfloat[]
{\includegraphics[width=.4\textwidth]{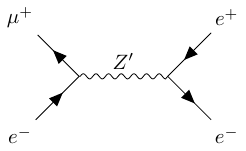}}
\hspace{10pt}
\subfloat[]
{\includegraphics[width=.4\textwidth]{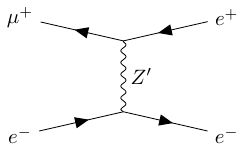}}
\caption{The Feynman diagrams of the process $\mu ^+ e^- \to e^+ e^-$: (a)~s-channel and (b)~t-channel.}
\label{fig:Feynman diagram}
\end{figure}

The signal cLFV processes studied in this paper are listed in Tab.~\ref{tab:physics process}. There are two diagrams for the $Z'$ mediated cLFV process $\mu ^+ e^- \to l^{+}l^{-}$, as shown in Fig.~\ref{fig:Feynman diagram} (taking $\mu ^+ e^- \to e^+ e^-$ as an example). In particular, the s-channel is not included in the processes $\mu ^+ e^- \to \mu^+ \tau^-$. In the simulation, all coupling strengths are considered as 1. And the mass of $Z'$ floats from 0.2~TeV to 5~TeV in the electron muon collision experiment, and within 0.50~GeV in the electron-target experiment with a muon beam.

While several background processes may occur on the collider and affect the signal that we are interested in, these processes include the standard model backgrounds divided by the number of final state particles and accidental background caused by particle mis-identification. Due to the suppression of phase space, we only consider the SM background with no more than 6 final state particles. In conclusion, the specific signals and their background processes are shown in Tab.~\ref{tab:physics process}.

\begin{table}[!t]
\centering
\caption{Specific processes of signal and SM background}
\label{tab:physics process}
\begin{tabular}{c c c c }
\toprule
\midrule
Signal Process&SM and accidental backgrounds.\\
\midrule
&$\mu ^+ e^- \to e^+ e^- \nu_e \Bar{\nu}_\mu$\\
\qquad$\mu ^+ e^- \to e^+ e^-$&$\mu ^+ e^- \to e^+ e^- \nu_e \Bar{\nu}_\mu \nu \Bar{\nu}$\\
&$\mu ^+ e^- \to \mu ^+ e^-$\\
\midrule
&$\mu ^+ e^- \to \mu^+ \mu^- \nu_e \Bar{\nu}_\mu$\\
\qquad$\mu ^+ e^- \to \mu^+ \mu^-$&$\mu ^+ e^- \to \mu^+ \mu^- \nu_e \Bar{\nu}_\mu \nu \Bar{\nu}$\\
&$\mu ^+ e^- \to \mu ^+ e^-$\\
\midrule
\qquad \multirow{2}*{$\mu ^+ e^- \to \mu^+ \tau^-$}&$\mu ^+ e^- \to \mu^+ \tau^- \nu_e \Bar{\nu}_\tau$\\
~&$\mu ^+ e^- \to \mu^+ \tau^- \nu_e \Bar{\nu}_\tau \nu \Bar{\nu}$\\
\midrule
\bottomrule
\end{tabular}
\end{table}

\subsection{Event generation and simulation}
\label{sec:simulation}

Cross section and Monte Carlo events for each signal and background are calculated and simulated by MadGraph5\_aMC@NLO(MG)~\cite{Madgraph} version 3.1.1, then showered and hadronized by Pythia8~\cite{Pythia}. Next, Delphes~\cite{Delphes} version 3.5.1 is utilized to simulate detector effects with the default configuration card for the detector at the muon collider.

In the Monte Carlo generation, some preliminary requirements are applied to remove the physically unreasonable events. In $e\mu$ collisions, the transverse momentum of charged leptons is required to satisfy $p_T >$  10~GeV and the absolute pseudo-rapidity of charged leptons $\lvert \eta \rvert >$ 2.5. While in the muon-beam electron-target simulation, the filtering criteria of $p_T$ and $\lvert \eta \rvert$ would be relaxed. Then in the detector simulation, the parameters such as particle detection efficiency are set according to the default detector configuration in the Delphes cards.

There are some differences between the $\tau$ simulation and $e/\mu$. For those final states with $\tau$ in Tab.~\ref{tab:physics process}, although they can go through any decay chains in reality, in this study we only consider the hadronic decay channels (about 60\% of the total decay). The $\tau$ tagging efficiency is 80\% with $p_T > 10$~GeV, as defined in Delphes card.

\section{Statistical analysis and sensitivity result}

\subsection{Asymmetric collision}

In this scenario we consider two kinds of asymmetric collision with electron and anti-muon beam: $E_e = 30$~GeV and $E_\mu = 1$~TeV ($\sqrt{s} = 346$~GeV), or $E_e = 200$~GeV and $E_\mu = 3$~TeV ($\sqrt{s} = 1.55$~TeV). The former is based on the proposal of $\mu$TRISTAN (the polarization of each beam is not considered), and the latter is a higher energy assumption according to other current beam designs.

\subsubsection{Background study}

\begin{figure}[htbp]
\centering
\subfloat[$\mu ^+ e^- \to W e^- \Bar{\nu}_\mu$]
{\includegraphics[width=.33\textwidth]{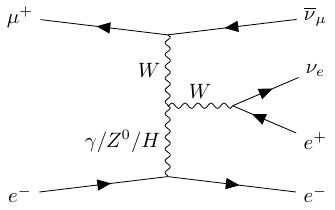}}
\subfloat[$\mu ^+ e^- \to Z^0(\gamma/H) \nu _e \Bar{\nu}_\mu$]
{\includegraphics[width=.33\textwidth]{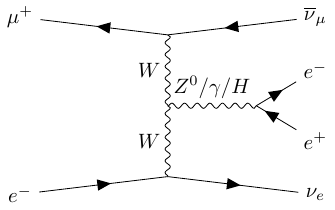}}
\subfloat[$\mu ^+ e^- \to e^+ e^- \nu _e \Bar{\nu}_\mu$]
{\includegraphics[width=.33\textwidth]{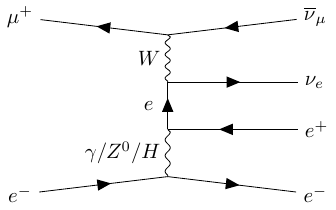}}
\caption{Typical diagrams for the SM background $\mu ^+ e^- \to l^+ l^- \nu \Bar{\nu}$, taking $\mu ^+ e^- \to e^+ e^- \nu \Bar{\nu}$ as an example. Especially, the diagram (b) does not appear in $\mu ^+ e^- \to \mu^+ \tau^- \nu \Bar{\nu}$.}
\label{fig:BKGFeynman}
\end{figure}

Unlike $e^+e^-$ or $\mu ^+ \mu ^-$ collisions, since the initial lepton flavor in $e\mu$ collision is non-zero, vast majority of the SM backgrounds are forbidden, for example, $l^+l^- \to W^+W^-$ and $l^+l^- \to \tau ^+ \tau ^-$. The typical diagrams for the SM background $\mu ^+ e^- \to l^+ l^- \nu \Bar{\nu}$ are shown in Fig.~\ref{fig:BKGFeynman}.

\begin{figure}[t]
\centering
\flushright
\subfloat[$\sqrt{s}=1.55$~TeV]
{\includegraphics[width=.49\textwidth]{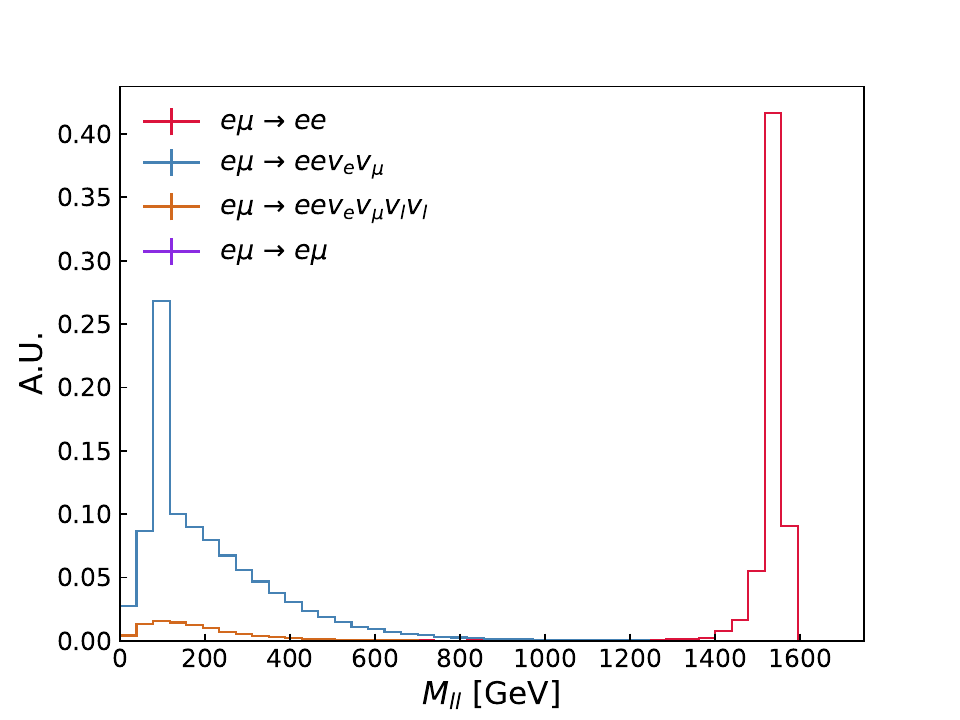}}
\subfloat[$\sqrt{s}=346$~GeV]
{\includegraphics[width=.49\textwidth]{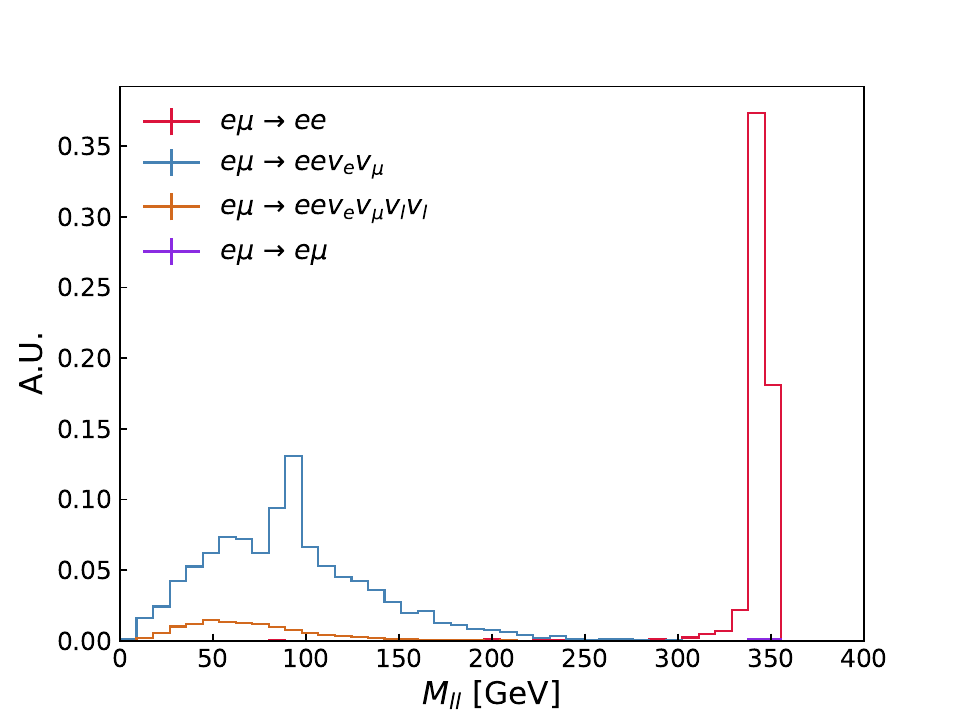}}
\vspace{0pt}
\subfloat[$\sqrt{s}=1.55$~TeV]
{\includegraphics[width=.49\textwidth]{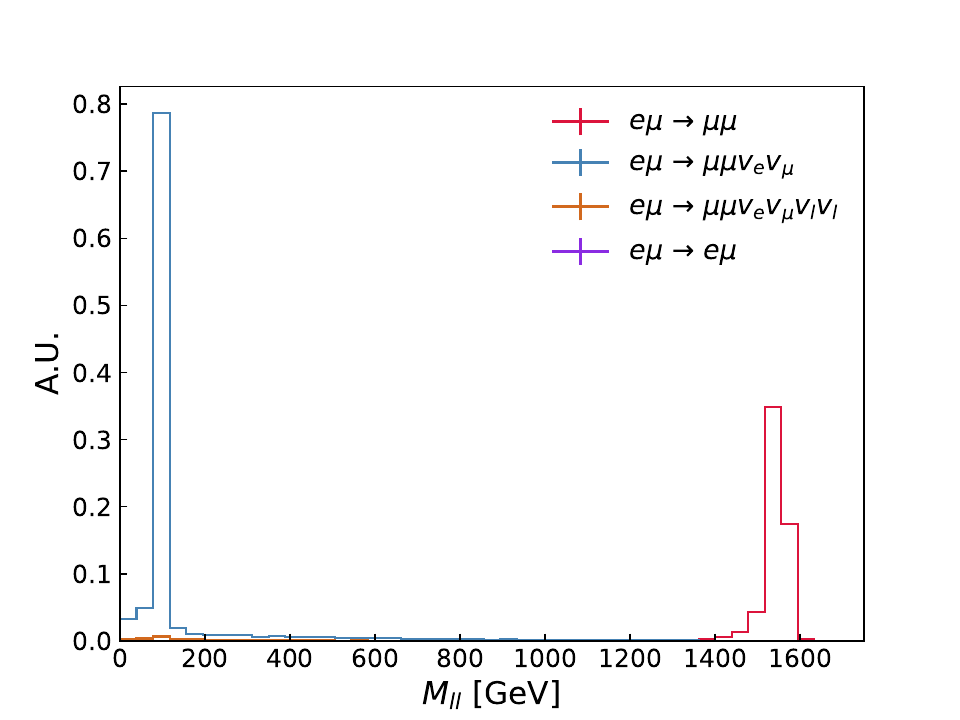}}
\subfloat[$\sqrt{s}=346$~GeV]
{\includegraphics[width=.49\textwidth]{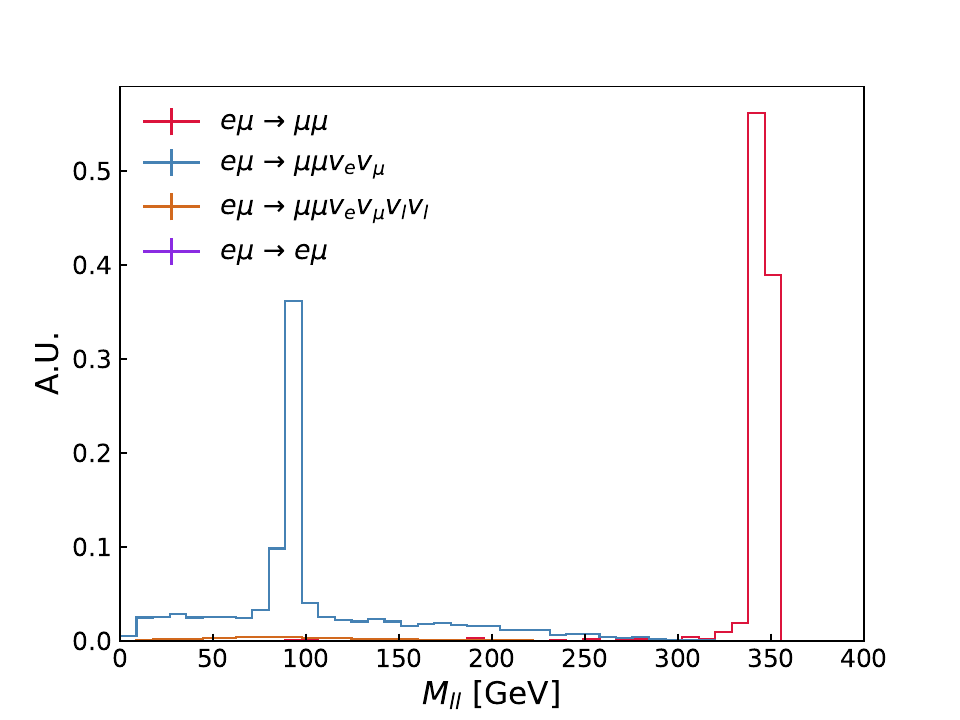}}
\vspace{0pt}
\subfloat[$\sqrt{s}=1.55$~TeV]
{\includegraphics[width=.49\textwidth]{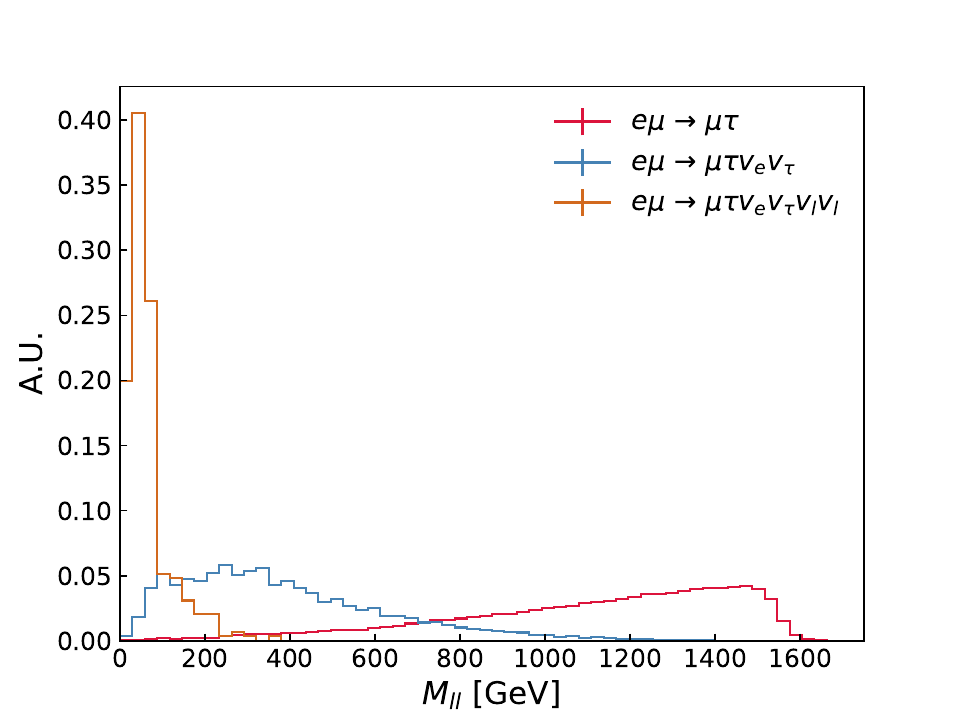}}
\subfloat[$\sqrt{s}=346$~GeV]
{\includegraphics[width=.49\textwidth]{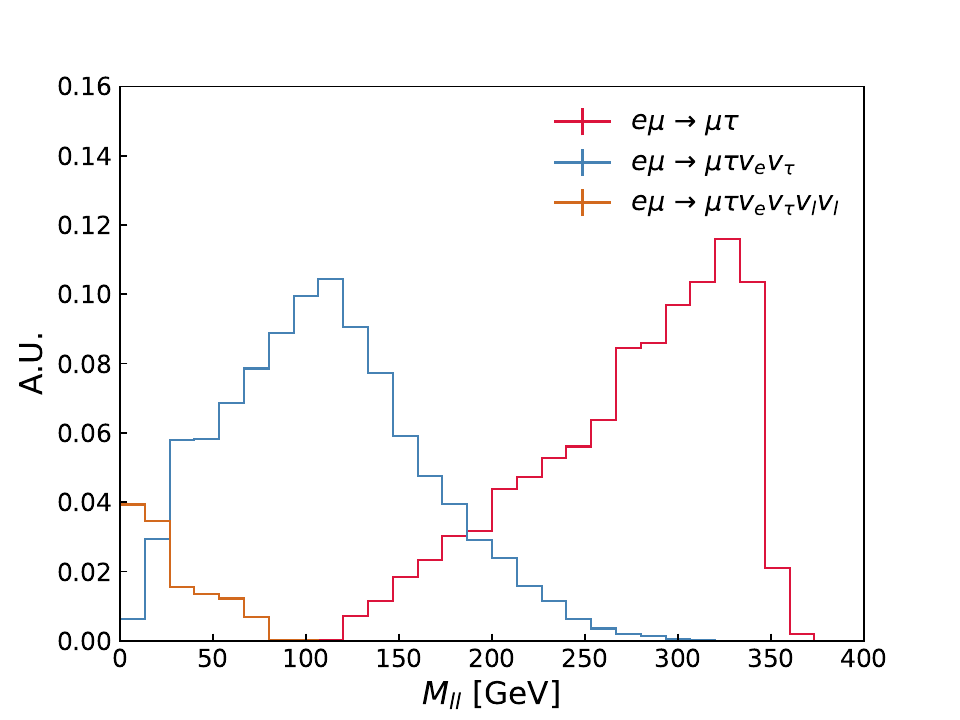}}
\caption{Distributions of the invariant mass reconstructed by the final state di-leptons, where the integral area of each background is determined by the cross section. Especially for the $e\mu$ scattering, it also includes the probability of particle mis-identification. Since the contribution of the background processes with six neutrinos is significantly lower than other kind of background, they will not be considered in the following analysis.}
\label{fig:mlldistribution1}
\end{figure}

After setting the preliminary requirements as mentioned in Sec.\ref{sec:simulation}, the signal candidate events should have the same charged leptons corresponding to the signal in the final state. According to the invariant mass distributions of the final state di-leptons as shown in Fig.~\ref{fig:mlldistribution1}, the remaining portion also exhibits significant kinematic differences from the signal processes, especially in $\mu ^+ e^- \to e^+ e^-$ or $\mu ^+ e^- \to \mu^+ \mu^-$ channel.

Only considering the interval near the center-of-mass energy, the SM background values are extremely low. And compared with a similar study on the different-signs electron or muon collider~\cite{Li:2023lin}, $e\mu$ collision has a cleaner signal window. Due to the low level of the SM background, we also investigated the accidental background caused by $e\mu$ mis-identification, where a final state muon is mis-identified as an electron, or vice versa. Typically it would let the $e\mu$ scattering process coming into the background of $\mu ^+ e^- \to e^+ e^-$ and $\mu ^+ e^- \to \mu^+ \mu^-$. The probability of $e\mu$ mis-identification is set as 10$^{-6}$. The invariant mass distribution of this process is extremely close to the signal. 

Other remaining backgrounds, such as those arising from $\mu - \pi$ mis-identification, are not included in this study, since it is difficult to produce a significant contribution to the background level when taking into account both the mis-identification probability and the kinematic differences from signal.

While for $\mu ^+ e^- \to \mu ^+ \tau ^-$, since the reconstruction of $\tau$ would inevitably result in a certain loss of energy, there is a considerable overlap in the signal and background distributions. The analysis will be optimized in the next section.

\subsubsection{Sensitivity result}

\label{sec:sens}

    Based on the distributions of the signal and background as shown in Fig.~\ref{fig:mlldistribution1}, we truncate the invariant mass to remove the events with the reconstructed invariant mass significantly deviating from center-of-mass energy. For the cLFV channels without $\tau$, the truncation points are set as 1.4~TeV for the 1.55~TeV collider, and 300~GeV for the $\mu$TRISTAN. While in $\tau$ channels, it will be determined by scanning and selecting the maximum figure of merit~(FOM) value of $S/(a/2+\sqrt{B})$, where $S$ is the number of signal events, $B$ is the weighted number of the background $\mu ^+ e^- \to \mu^+ \tau^- \nu_e \Bar{\nu}_\tau$ and $a$ is the significance value, which is usually set as 3 for new physics discoveries~\cite{Punzi:2003bu}. The weight for each process is defined by $n_x = \sigma _x L/N$, where $\sigma _x$ is the cross section, $L$ is the luminosity and $N$ is the generated event number, which is 200,000 for each process. When $L =0.3$~ab$^{-1}$, the truncation point results are 1.27~TeV and 280~GeV for the two collider.

\begin{figure}[!t]
    \centering
    \includegraphics[width=.7\textwidth]{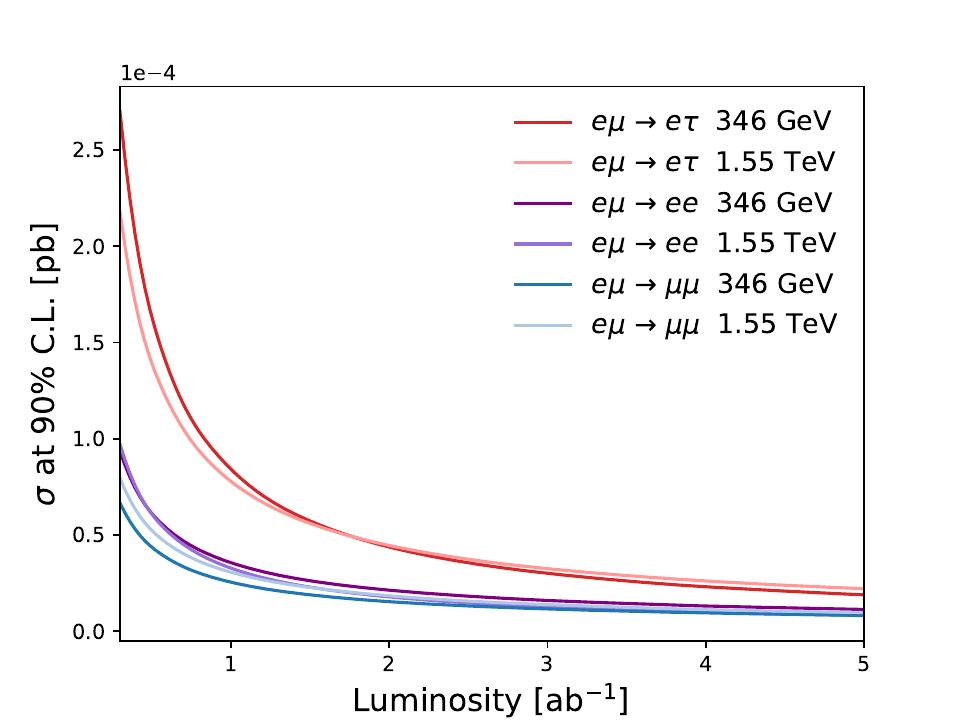}
    \caption{Cross section of each cLFV signal process of 90\% C.L. exclusion relative to the luminosity of $e\mu$ collider.}
    \label{fig:cross}
\end{figure}

Then the binned histograms of leptons $p_T$ distributions are utilized for the statistical analysis. The test statistics $Z_i$ is calculated by $Z_i := 2[n_i - b_i +b_i \ln{(n_i / b_i)}]$ for 90\% confidence level (C.L.) exclusion, where $i$ is the index of each bin, $n$ is the weighted number of observed events including signal and background, and $b$ is the weighted number of background~\cite{Cowan:2010js}. Then the total $Z=\Sigma _{i} Z_i$ would be subject to a $\chi ^2$ distribution. The degree of freedom is defined by the number of bins. By iteration, we can obtain the signal cross section of 90\% C.L. exclusion, as shown in Fig.~\ref{fig:cross}. 

\begin{figure}
\centering
\subfloat[$\mu ^+ e^- \to e^+ e^-$]
{\includegraphics[width=.5\textwidth]{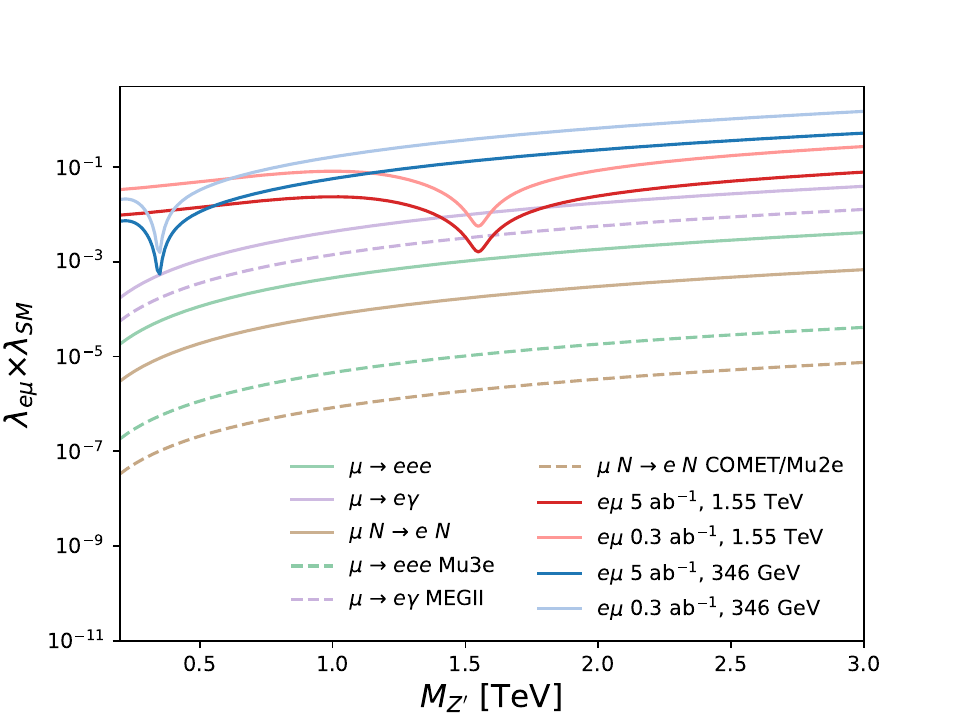}\label{fig:lambda1a}}
\subfloat[$\mu ^+ e^- \to \mu^+ \tau^-$]
{\includegraphics[width=.5\textwidth]{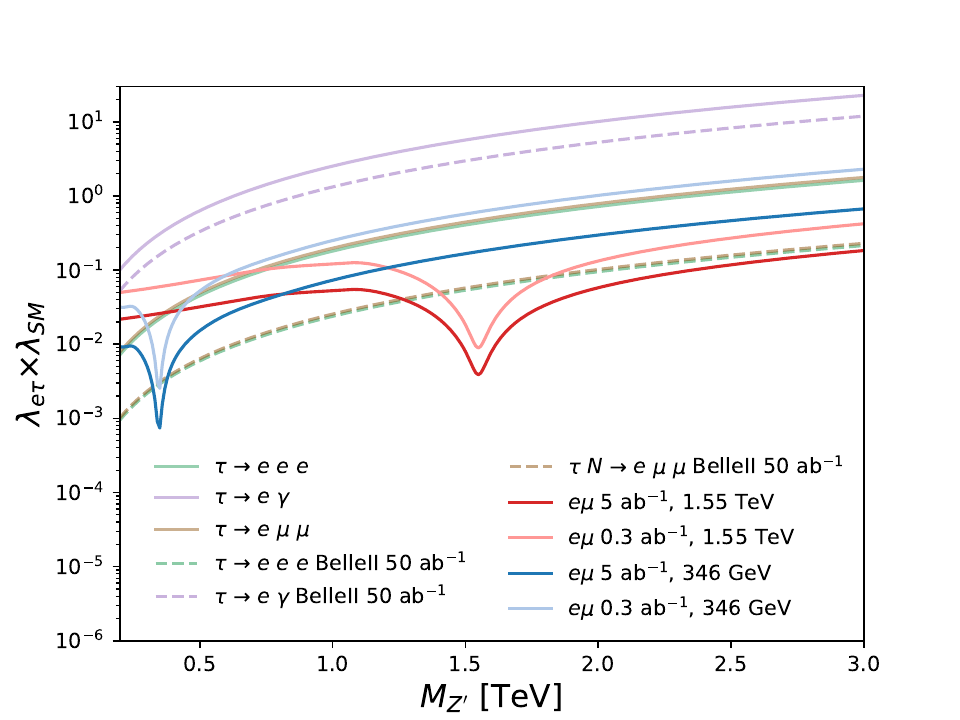}\label{fig:lambda1b}}
\caption{The 90\% C.L. upper limit of $\lambda_{e \mu} \times \lambda_{\rm SM}$ (a) and $\lambda_{e\tau} \times \lambda_{\rm SM}$ (b). The curves are graphed with respect to $M_{Z^{'}}$, representing the limits of the cross section times branching ratio. Additionally, exclusion lines from both present low-energy experiments (shown as solid lines in purple, green and brown) and future experiments (shown as dashed lines in purple, green and brown) are included in the plot.}
\label{fig:lambda1}
\end{figure}

Then based on MadGraph calculation we can get the corresponding value of $\lambda _{ll'} \times \lambda _{\rm SM}$, as shown in Fig.~\ref{fig:lambda1}. Currently the coupling $\lambda _{ij}$ are considered as positive real. Then the $\lambda _{e\mu}$ and $\lambda _{e\tau}$ results are calculated at two energy points, and the luminosity is 0.3~ab$^{-1}$ or 5~ab$^{-1}$. Several current limits and prospective limits are also included to compare with our results. The constraints of branching ratio from other experiments are concluded in Tab.~\ref{tab:limits}. It is sensible that the limits given by other experiments in $\tau$ channels are much more conservative than in $e\mu$ channels. But in our collider study, they may be similar since the main influencing factor here is the signal cross section while the cross sections of different signals are quite similar.

\begin{table}[!t]
\centering
\caption{Summary of the current and prospective limits from other experiments at 90\% C.L.}
\label{tab:limits}
\begin{tabular}{c c c c }
\toprule
\midrule
\multirow{2}*{Coupling}&\multirow{2}*{Channel}&\multicolumn{2}{c}{Constraint of branching ratio}~\\
~&~&Current&Prospective\\
\midrule
&$\mu N \to e N$&$6.1 \times 10^{-13}$~\cite{SINDRUMII:1998mwd}&$3.0 \times 10^{-17}$~\cite{COMET:2018auw}\\
$\lambda _{e\mu}$&$\mu \to e e e$&$1.0 \times 10^{-13}$~\cite{SINDRUM:1987nra}&$1.0 \times 10^{-16}$~\cite{Mu3e:2020gyw}\\
&$\mu \to e \gamma$&$4.2 \times 10^{-13}$~\cite{MEG:2016leq}&$6.0 \times 10^{-14}$~\cite{MEGII:2018kmf}\\
\midrule
&$\tau \to e \gamma$&$3.3 \times 10^{-8}$~\cite{Hayasaka:2010np}&$9.0 \times 10^{-9}$~\cite{Banerjee:2022vdd}\\
$\lambda _{e\tau}$&$\tau \to \mu \mu e$&$2.7 \times 10^{-8}$~\cite{Hayasaka:2010np}&$4.5 \times 10^{-10}$~\cite{Banerjee:2022vdd}\\
&$\tau \to e e e$&$2.7 \times 10^{-8}$~\cite{Hayasaka:2010np}&$4.7 \times 10^{-10}$~\cite{Banerjee:2022vdd}\\
\midrule
\bottomrule
\end{tabular}
\end{table}

In the $\lambda _{e\mu}$ channel, the strictest constraint comes from the results of $\mu - e$ conversion, and our results have weaker competitiveness among those high intensity muon-based experiments. While in the $\lambda _{e\tau}$ channel, current limits of $\tau \to e e e$ and $\tau \to e \mu \mu$ perform the best across the entire interval, but the orders of magnitude are much lower than the results of $\lambda _{e\mu}$. Compared with our results, the constraints of 1.55~TeV $e\mu$ collider are more stringent than other existing results when $M_{Z'} > 0.5$~TeV, and for the 346~GeV collider it is about the entire $M_{Z'}$ interval. Even compared with the prospective limits of those processes on Belle II, our results still have certain advantages around the resonance region.

On the other side, comparing the results of $e\mu$ collision and $e^+e^-$ or $\mu ^+ \mu ^-$ collisions~\cite{Li:2023lin} based on the same luminosity, the results of $e\mu$ are better than $\mu ^+ \mu ^-$ by an order of magnitude at their respective resonance points. But it cannot be ignored that the center-of-mass energy of $e\mu$ is relatively low, and the results of other experiments are more advantageous in the low-energy region. Furthermore, it is difficult for $e\mu$ collider to reach the same luminosity as that of $\mu ^+ \mu ^-$ collider in reality. Therefore, the advantages of these two collision scenarios still need to be considered based on more practical factors.

\begin{figure}
    \centering
    \includegraphics[width=.65\textwidth]{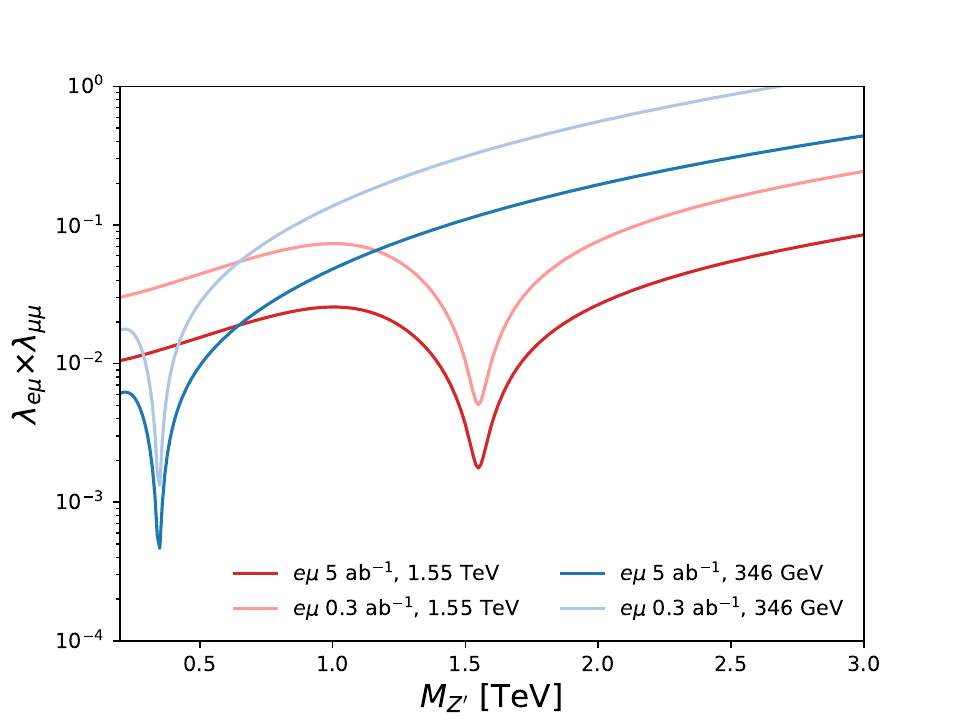}
    \caption{Sensitivity of the coupling strength $\lambda _{\mu\mu} \times \lambda_{e\mu}$ from $\mu ^+ e^- \to \mu^+ \mu^-$ channel. Current and prospective results of $\mu \to e\gamma$ are include for comparison, assuming that it is entirely composed of $\lambda _{e\mu} \times \lambda _{\mu\mu}$.}
    \label{fig:lambda1c}
\end{figure}

To be precise, what we are comparing in Fig.~\ref{fig:lambda1a} and Fig.~\ref{fig:lambda1b} is the coupling $\lambda _{e\mu} (\lambda _{e\tau}) \times \lambda _{\rm SM}$. Then based on the assumption that $Z'$ has the same coupling structures and strengths as those of the standard model $Z^0$ mentioned in Sec.~\ref{sec:zpmodel}, which is $\lambda _{\rm SM} = 1$, we can naturally obtain the estimation of $\lambda _{e\mu}$ or $\lambda _{e\tau}$. However, strictly speaking, this requirement may not exist in more universal models, and we need to consider the biases of standard model couplings. In this way, each process may give a product of different coupling strength, for example $\lambda _{e\mu} \times \lambda _{ee}$ ($\mu \to eee$) and $\lambda _{e\mu} \times \lambda _{ll}$ ($\mu \to e\gamma$), where $l$ represents $e$ or $\mu$. Specially, we can obtain the measurement of the coupling $\lambda _{e\mu} \times \lambda _{\mu\mu}$ through $\mu ^+ e^- \to \mu ^+ \mu ^-$ based on the $e\mu$ collider, as shown in Fig.~\ref{fig:lambda1c}, indicating that it has similar sensitivity as $\mu ^+ e^- \to e^+ e^-$. While for the current high intensity experiments, this coupling only appears in $\mu \to e\gamma$ together with $\lambda _{e\mu} \times \lambda _{ee}$ according to Ref.~\cite{Langacker:2000ju}, as 
\begin{equation}
    \Gamma(\mu \to e\gamma)  = \frac{\alpha G_F^2m^3_{\mu}M^4_{Z}}{4\pi ^4M^4_{Z'}} \left[\sin ^2 \theta_W  \left(\sin ^2 \theta_W-\frac{1}{2}\right)\right]^2 \left(\lambda_{ee}\lambda_{e\mu}m_{e} + \lambda_{e\mu}\lambda_{\mu\mu}m_{\mu} \right)^2,
        \label{eq:muegwidth}
\end{equation}
where $\alpha$ is the fine structure constant, $G_F$ is the fermi constant, $\theta _W$ is the weak mixing angle, $M_{Z}$, $M_{Z'}$, $m_\mu$ and $m_e$ is the mass of $Z^0$, $Z'$, $\mu$ and $e$, respectively.

\begin{figure}
    \centering
    \includegraphics[width=.65\textwidth]{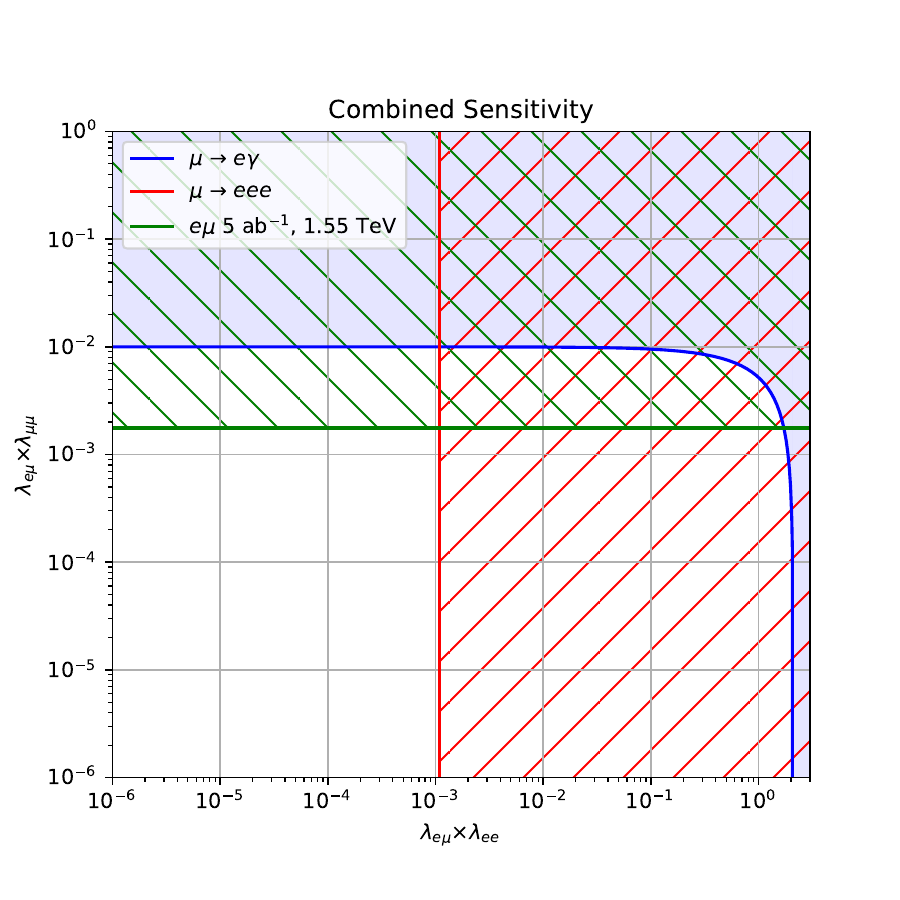}
    \caption{Sensitivity of the combined coupling strength at the resonance point of the 1.55~TeV $e\mu$ collider. The results of $\mu \to eee$ and $\mu \to e\gamma$ are calculated by the current limits. Each shaded area serves as the exclusion zone for the corresponding process.\label{fig:lambda1d}}
\end{figure}

Since the contribution of $\lambda _{e\mu} \times \lambda _{ee}$ is suppressed by the ratio of lepton masses, we can assume that this process is entirely composed of $\lambda _{e\mu} \times \lambda _{\mu\mu}$ and is comparable with our result as shown in Fig.~\ref{fig:lambda1c}. Meanwhile, if considering the rigorous contributions, it set the joint limitations on these two coupling strength, as shown in Fig.~\ref{fig:lambda1d}.

\subsection{Electron-target experiment with a muon beam}

\begin{figure}[H]
\centering
\flushright
\subfloat[$\mu ^+ e^- \to e^+ e^-$]
{\includegraphics[width=.5\textwidth]{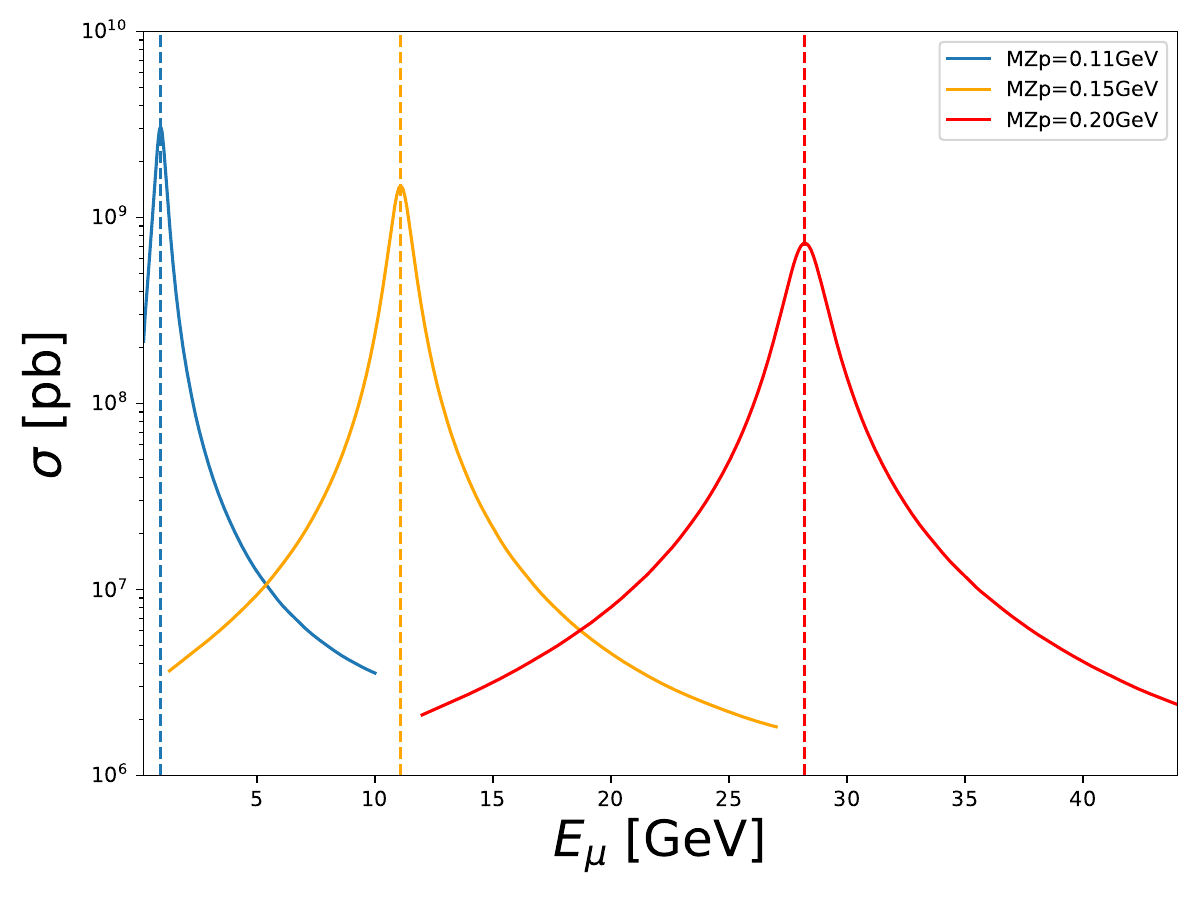}}
\subfloat[$\mu ^+ e^- \to \mu ^+ \mu ^-$]
{\includegraphics[width=.5\textwidth]{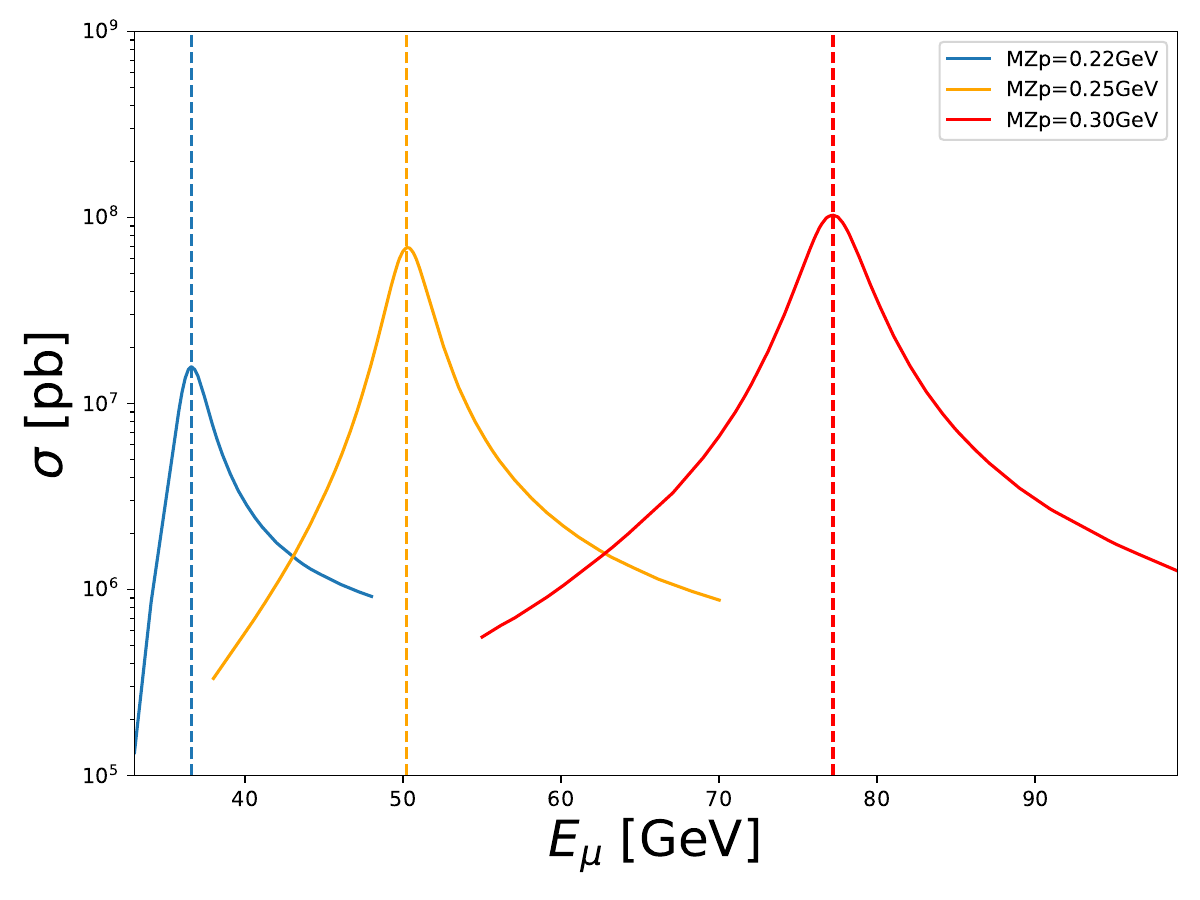}}
\vspace{-5pt}
\caption{Cross section for resonant production of the process $\mu ^+ e^- \to e^+ e^-$ and $\mu ^+ e^- \to \mu ^+ \mu ^-$with different $M_{Z^{'}}$.}
\label{fig:resonance}
\end{figure}

 Now we will focus on the low-energy region, where we conduct Monte Carlo simulations to investigate cLFV processes involving $\mu ^+ e^- \to e^+ e^-$ and $\mu ^+ e^- \to \mu ^+ \mu ^-$ in the electron-target experiment with a muon beam. It is essential to highlight that the masses of muon and electron cannot be disregarded, imposing a lower limit on the incident muon beam energy, equivalent to its mass. Additionally, the target experiment energy $E_{cm}=\sqrt{2E_{\mu}m_e + {m_{\mu}}^2 +{m_e}^2}$ possesses its own lower-energy threshold. Consequently, in our simulations, we vary the $M_{Z^{'}}$ in three distinct sets for each process as shown in Tab.~\ref{tab:resonance}, ensuring that the muon energy is scanned as close to the lower energy limit as feasible. The outcomes are shown in Fig.~\ref{fig:resonance}. Remarkably, a pronounced resonance in the target energy is observed in proximity to $M_{Z^{'}}$.

 \begin{table}[t]
 \caption{Resonant collision energy of process $\mu ^+ e^- \to e^+ e^-$ and $\mu ^+ e^- \to \mu ^+ \mu ^-$ with different $M_{Z^{'}}$.}
 \centering
	\begin{tabular}{c c c c c }
		\toprule\midrule
		Process & $M_{Z^{'}}$ / $\si{\GeV}$ & $E_{\mu}$ / $\si{\GeV}$ & 
		$E_{e}$ / $\si{\MeV}$  &  $E_{cm}$ / $\si{\GeV}$\\
		\midrule
		\multirow{3}{*}{$\mu ^+ e^- \to e^+ e^-$} & 0.11 & 0.93 & 0.511 & 0.1101 \\
		& 0.15 & 11.1 & 0.511 & 0.1501 \\
		& 0.20 & 28.2 & 0.511 & 0.1996 \\
		\midrule
		\multirow{3}{*}{$\mu ^+ e^- \to \mu ^+ \mu ^-$} & 0.22 & 33.6 & 0.511 & 0.2200 \\
		& 0.25 & 50.2 & 0.511 & 0.2499 \\
		& 0.30 & 77.2 & 0.511 & 0.2998 \\
		\midrule\bottomrule
	\end{tabular}	
\label{tab:resonance}
\end{table}

In this scenario, we conduct a background-free experimental estimation. Due to $\sigma \propto (\lambda _{e \mu}\times \lambda _{ll})^2$ and event rate $R = \frac{dN}{dt} \cdot n \cdot dx \cdot \sigma$, where $\frac{dN}{dt}$ denotes the muon production rate, $n$ denotes the electron number density of the target material, $dx$ denotes the thickness of the target, and $\sigma$ denotes the reaction cross section, coupling limits estimates can be made based on the reaction cross-section. As a rough estimate, assuming a 10 cm thick lead target, the incidence rate is about $\frac{dN}{dt}\sim  10^6 \rm{s^{-1}}$, the electron number density of lead is about $n \sim 10^{24} \rm{cm^{-3}}$, and $1 \mathrm{y} \sim 10^7 \si{\s}$. From this estimation, we can obtain the 90\% C.L. exclusion lines on the couplings $\lambda_{ee}$ and $\lambda_{\mu \mu}$ products the diagonal coupling $\lambda_{e \mu}$. 

\begin{figure}[H]
\centering
\flushright
\subfloat[$\mu ^+ e^- \to e^+ e^-$]
{\includegraphics[width=.5\textwidth]{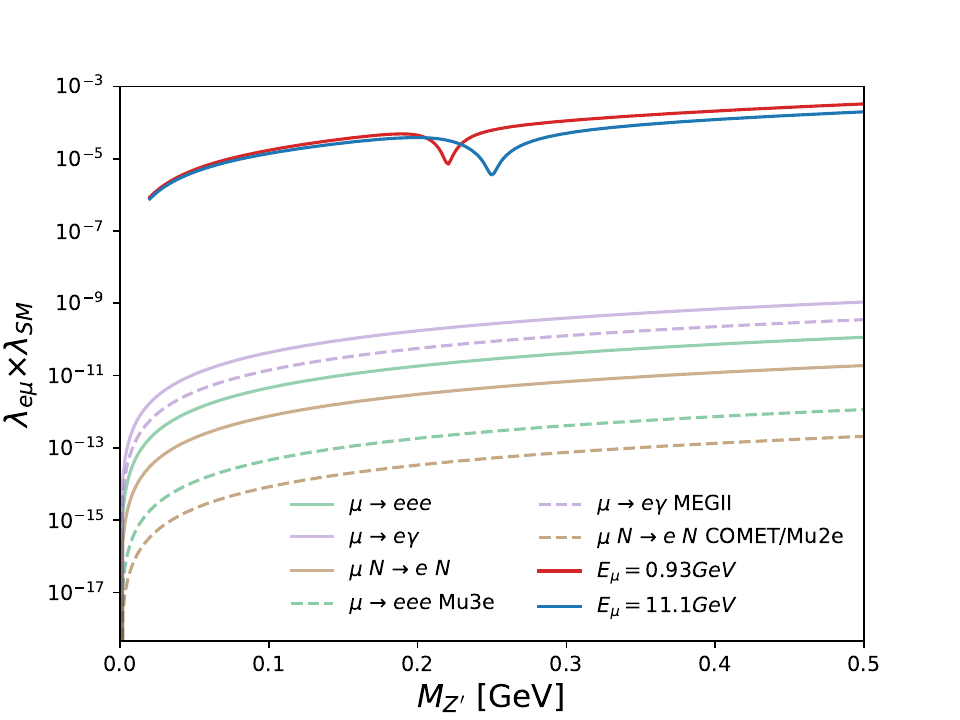}\label{fig:3.2.1}}
\subfloat[$\mu ^+ e^- \to \mu ^+ \mu ^-$]
{\includegraphics[width=.5\textwidth]{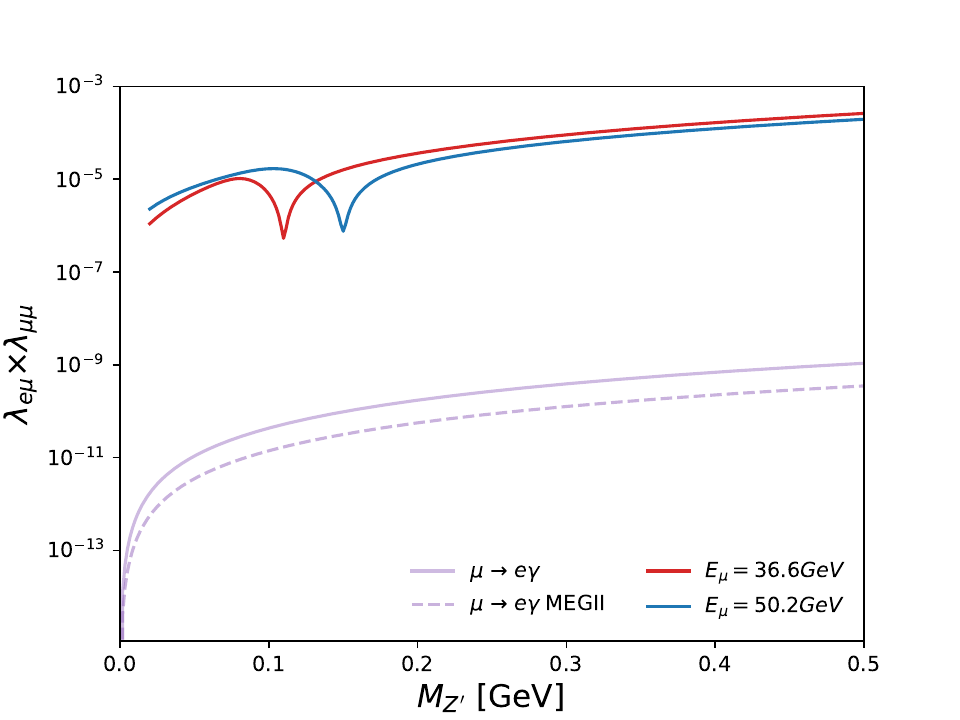}\label{fig:3.2.2}}
\vspace{-5pt}
\caption{The 90\% C.L. upper limit of $\lambda_{e \mu} \times \lambda_{\rm SM}$ (a) and $\lambda_{e\mu} \times \lambda_{\mu \mu}$ (b). The curves are graphed with respect to $M_{Z^{'}}$, representing the limits of the cross section times branching ratio. Additionally, exclusion lines from both present low-energy experiments (shown as solid lines in purple, green and brown) and future experiments (shown as dashed lines in purple, green and brown) are included in the plot.}
\end{figure}

Similarly, the current and prospective limits from low-energy experiments are converted to the coupling limits on $\lambda_{e \mu}\times\lambda_{\rm SM}$ to compare with our results. As shown in Fig.~\ref{fig:3.2.1}, compared with the existing experimental results and expected experimental results, the restrictions given here do not have obvious advantages. However, considering the $\mu \mu$ channel, if the coupling between $Z'$ and $\mu \mu$ is much stronger than $e\mu$, that is, $\lambda _{e\mu}\ll  \lambda _{\mu\mu}$, and assuming that the coupling strengths of other channels are all set to 0, then $Z'$ decay mostly into  $\mu \mu$, and the production cross section depends only on $\lambda _{e\mu}$. Accordingly, the resulted limit on $\lambda _{e\mu}$, compared with $\mu \to e \gamma$, can gain advantage of several orders of magnitudes.  At the same time, if the data accumulation time of the muon beam experiment is considered to be 10 years, our estimated results will drop by another order of magnitude, which will improve the simulation results to a certain extent. However, it is worth noting that under the above estimation of the coupling strengths, $\Gamma _{Z}$ will be reduced, which means that $Z'$ will not be on resonance easily.

Additionally, since the result of $\mu \to e \gamma$ is related to the combination of $m_e \lambda_{e \mu} \lambda_{e e}$ and $m_\mu \lambda_{\mu \mu} \lambda_{e \mu}$ as Eq.~\ref{eq:muegwidth}, their inference may be non-ignorable. While our detection is exclusively sensitive to the $m_e \lambda_{e \mu} \lambda_{e e}$ term or the $m_\mu \lambda_{\mu \mu} \lambda_{e \mu}$ term, depending on the channel we choose.

\section{Conclusion}

With the continuous development of muon technology, in addition to building high-energy muon collider, there is also certain research prospect for the $e\mu$ interaction. In this work, we investigate the cLFV processes propagated by a massive neutral gauge boson ($Z'$) in $e\mu$ collision and electron-target experiment with a muon beam, in order to explore the potential of $e\mu$ interactions in new physics searches. Using the MadGraph5\_aMC@NLO, Pythia8 and Delphes software, we conduct the simulation studies on the cLFV processes $\mu ^+ e^- \to e^+ e^-$, $\mu ^+ e^- \to \mu^+ \mu^-$ and $\mu ^+ e^- \to \mu^+ \tau^-$. We then provide the coupling strength $\lambda _{e\mu}$ and $\lambda _{e\tau}$ at the 90\% C.L. for different mass of ${Z'}$. By comparing the sensitivity results with the limits from the current and prospective experiments, it is shown that $e\mu$ interactions have certain research advantages in $\tau$ channel for the heavy $Z'$. Furthermore, based on the universal assumption, a unique process corresponding to coupling $\lambda_{e\mu}\times\lambda_{\mu\mu}$ can be measured, while this coupling combination can only appear in $\mu \to e\gamma$ channel among the current experiments.

\acknowledgments

This work is supported in part by the National Natural Science Foundation of China under Grant No. 12150005, 12075004, 12175321, 12061141003; the National Key Research and Development Program of China under Grant No. 2018YFA0403900; National College Students Innovation and Entrepreneurship Training Program, Sun Yat-sen University.


\bibliographystyle{JHEP}
\bibliography{main.bib}

\providecommand{\href}[2]{#2}\begingroup\raggedright\begin{thebibliography}{10}

\bibitem{Apollinari:2017lan}
G.~Apollinari, I.~B\'ejar~Alonso, O.~Br\"uning, P.~Fessia, M.~Lamont, L.~Rossi et~al., eds., \emph{{High-Luminosity Large Hadron Collider (HL-LHC)}: {Technical Design Report V. 0.1}}, .

\bibitem{FCC:2018byv}
{\scshape FCC} collaboration, \emph{{FCC Physics Opportunities}: {Future Circular Collider Conceptual Design Report Volume 1}}, \href{https://doi.org/10.1140/epjc/s10052-019-6904-3}{\emph{Eur. Phys. J. C} {\bfseries 79} (2019) 474}.

\bibitem{FCC:2018evy}
{\scshape FCC} collaboration, \emph{{FCC-ee: The Lepton Collider}: {Future Circular Collider Conceptual Design Report Volume 2}}, \href{https://doi.org/10.1140/epjst/e2019-900045-4}{\emph{Eur. Phys. J. ST} {\bfseries 228} (2019) 261}.

\bibitem{FCC:2018vvp}
{\scshape FCC} collaboration, \emph{{FCC-hh: The Hadron Collider}: {Future Circular Collider Conceptual Design Report Volume 3}}, \href{https://doi.org/10.1140/epjst/e2019-900087-0}{\emph{Eur. Phys. J. ST} {\bfseries 228} (2019) 755}.

\bibitem{CEPCStudyGroup:2018rmc}
{\scshape CEPC Study Group} collaboration, \emph{{CEPC Conceptual Design Report: Volume 1 - Accelerator}},  \href{https://arxiv.org/abs/1809.00285}{{\ttfamily 1809.00285}}.

\bibitem{CEPCStudyGroup:2018ghi}
{\scshape CEPC Study Group} collaboration, \emph{{CEPC Conceptual Design Report: Volume 2 - Physics \& Detector}},  \href{https://arxiv.org/abs/1811.10545}{{\ttfamily 1811.10545}}.

\bibitem{Aime:2022flm}
C.~Aime et~al., \emph{{Muon Collider Physics Summary}},  \href{https://arxiv.org/abs/2203.07256}{{\ttfamily 2203.07256}}.

\bibitem{Barger:1997dv}
V.D.~Barger, S.~Pakvasa and X.~Tata, \emph{{Are e mu colliders interesting?}}, \href{https://doi.org/10.1016/S0370-2693(97)01234-3}{\emph{Phys. Lett. B} {\bfseries 415} (1997) 200} [\href{https://arxiv.org/abs/hep-ph/9709265}{{\ttfamily hep-ph/9709265}}].

\bibitem{Choi:1997bm}
S.Y.~Choi, C.S.~Kim, Y.J.~Kwon and S.-H.~Lee, \emph{{High-energy FCNC search through e mu colliders}}, \href{https://doi.org/10.1103/PhysRevD.57.7023}{\emph{Phys. Rev. D} {\bfseries 57} (1998) 7023} [\href{https://arxiv.org/abs/hep-ph/9707483}{{\ttfamily hep-ph/9707483}}].

\bibitem{Montero:1998sv}
J.C.~Montero, V.~Pleitez and M.C.~Rodriguez, \emph{{Left-right asymmetries in polarized e - mu scattering}}, \href{https://doi.org/10.1103/PhysRevD.58.097505}{\emph{Phys. Rev. D} {\bfseries 58} (1998) 097505} [\href{https://arxiv.org/abs/hep-ph/9803450}{{\ttfamily hep-ph/9803450}}].

\bibitem{Cvetic:1999fk}
G.~Cvetic and C.S.~Kim, \emph{{Heavy Majorana neutrino production at electron - muon colliders}}, \href{https://doi.org/10.1016/S0370-2693(99)00848-5}{\emph{Phys. Lett. B} {\bfseries 461} (1999) 248} [\href{https://arxiv.org/abs/hep-ph/9906253}{{\ttfamily hep-ph/9906253}}].

\bibitem{Almeida:2000qd}
F.M.L.~Almeida, Jr., Y.~do~Amaral~Coutinho, J.A.~Martins~Simoes and M.A.B.~Vale, do., \emph{{Single neutral heavy lepton production at electron muon colliders}}, \href{https://doi.org/10.1016/S0370-2693(00)01195-3}{\emph{Phys. Lett. B} {\bfseries 494} (2000) 273} [\href{https://arxiv.org/abs/hep-ph/0008231}{{\ttfamily hep-ph/0008231}}].

\bibitem{Singhal:2007hw}
J.K.~Singhal, S.~Singh and A.K.~Nagawat, \emph{{Possible exotic neutrino signature in electron muon collisions}},  \href{https://arxiv.org/abs/hep-ph/0703136}{{\ttfamily hep-ph/0703136}}.

\bibitem{Bouzas:2023vba}
A.O.~Bouzas and F.~Larios, \emph{{An electron-muon collider: what can be probed with it?}}, \href{https://doi.org/10.31349/SuplRevMexFis.4.021128}{\emph{Rev. Mex. Fis. Suppl.} {\bfseries 4} (2023) 021128}.

\bibitem{Bossi:2020yne}
F.~Bossi and P.~Ciafaloni, \emph{{Lepton Flavor Violation at muon-electron colliders}}, \href{https://doi.org/10.1007/JHEP10(2020)033}{\emph{JHEP} {\bfseries 10} (2020) 033} [\href{https://arxiv.org/abs/2003.03997}{{\ttfamily 2003.03997}}].

\bibitem{Lu:2020dkx}
M.~Lu, A.M.~Levin, C.~Li, A.~Agapitos, Q.~Li, F.~Meng et~al., \emph{{The physics case for an electron-muon collider}}, \href{https://doi.org/10.1155/2021/6693618}{\emph{Adv. High Energy Phys.} {\bfseries 2021} (2021) 6693618} [\href{https://arxiv.org/abs/2010.15144}{{\ttfamily 2010.15144}}].

\bibitem{Bouzas:2021sif}
A.O.~Bouzas and F.~Larios, \emph{{Two-to-Two Processes at an Electron-Muon Collider}}, \href{https://doi.org/10.1155/2022/3603613}{\emph{Adv. High Energy Phys.} {\bfseries 2022} (2022) 3603613} [\href{https://arxiv.org/abs/2109.02769}{{\ttfamily 2109.02769}}].

\bibitem{Hamada:2022mua}
Y.~Hamada, R.~Kitano, R.~Matsudo, H.~Takaura and M.~Yoshida, \emph{{$\mu$TRISTAN}}, \href{https://doi.org/10.1093/ptep/ptac059}{\emph{PTEP} {\bfseries 2022} (2022) 053B02} [\href{https://arxiv.org/abs/2201.06664}{{\ttfamily 2201.06664}}].

\bibitem{Akturk:2024jbl}
D.~Akturk, B.~Dagli and S.~Sultansoy, \emph{{Muon Ring and FCC-ee / CEPC Based Antimuon-Electron Colliders}},  \href{https://arxiv.org/abs/2403.17034}{{\ttfamily 2403.17034}}.

\bibitem{Das:2024gfg}
A.~Das and Y.~Orikasa, \emph{{Z' induced forward dominant processes in \ensuremath{\mu}TRISTAN experiment}}, \href{https://doi.org/10.1016/j.physletb.2024.138577}{\emph{Phys. Lett. B} {\bfseries 851} (2024) 138577} [\href{https://arxiv.org/abs/2401.00696}{{\ttfamily 2401.00696}}].

\bibitem{Lichtenstein:2023iut}
G.~Lichtenstein, M.A.~Schmidt, G.~Valencia and R.R.~Volkas, \emph{{Complementarity of $\mu$TRISTAN and Belle II in searches for charged-lepton flavour violation}}, \href{https://doi.org/10.1016/j.physletb.2023.138144}{\emph{Phys. Lett. B} {\bfseries 845} (2023) 138144} [\href{https://arxiv.org/abs/2307.11369}{{\ttfamily 2307.11369}}].

\bibitem{Hamada:2022uyn}
Y.~Hamada, R.~Kitano, R.~Matsudo and H.~Takaura, \emph{{Precision \ensuremath{\mu}+\ensuremath{\mu}+ and \ensuremath{\mu}+e\ensuremath{-} elastic scatterings}}, \href{https://doi.org/10.1093/ptep/ptac174}{\emph{PTEP} {\bfseries 2023} (2023) 013B07} [\href{https://arxiv.org/abs/2210.11083}{{\ttfamily 2210.11083}}].

\bibitem{kahn2018m3}
Y.~Kahn, G.~Krnjaic, N.~Tran and A.~Whitbeck, \emph{{M$^{3}$: a new muon missing momentum experiment to probe (g \ensuremath{-} 2)$_{\mu}$ and dark matter at Fermilab}}, \href{https://doi.org/10.1007/JHEP09(2018)153}{\emph{JHEP} {\bfseries 09} (2018) 153} [\href{https://arxiv.org/abs/1804.03144}{{\ttfamily 1804.03144}}].

\bibitem{galon2020searching}
I.~Galon, E.~Kajamovitz, D.~Shih, Y.~Soreq and S.~Tarem, \emph{{Searching for muonic forces with the ATLAS detector}}, \href{https://doi.org/10.1103/PhysRevD.101.011701}{\emph{Phys. Rev. D} {\bfseries 101} (2020) 011701} [\href{https://arxiv.org/abs/1906.09272}{{\ttfamily 1906.09272}}].

\bibitem{Barbier:2004ez}
R.~Barbier et~al., \emph{{R-parity violating supersymmetry}}, \href{https://doi.org/10.1016/j.physrep.2005.08.006}{\emph{Phys. Rept.} {\bfseries 420} (2005) 1} [\href{https://arxiv.org/abs/hep-ph/0406039}{{\ttfamily hep-ph/0406039}}].

\bibitem{Dorsner:2016wpm}
I.~Dor\v{s}ner, S.~Fajfer, A.~Greljo, J.F.~Kamenik and N.~Ko\v{s}nik, \emph{{Physics of leptoquarks in precision experiments and at particle colliders}}, \href{https://doi.org/10.1016/j.physrep.2016.06.001}{\emph{Phys. Rept.} {\bfseries 641} (2016) 1} [\href{https://arxiv.org/abs/1603.04993}{{\ttfamily 1603.04993}}].

\bibitem{Branco:2011iw}
G.C.~Branco, P.M.~Ferreira, L.~Lavoura, M.N.~Rebelo, M.~Sher and J.P.~Silva, \emph{{Theory and phenomenology of two-Higgs-doublet models}}, \href{https://doi.org/10.1016/j.physrep.2012.02.002}{\emph{Phys. Rept.} {\bfseries 516} (2012) 1} [\href{https://arxiv.org/abs/1106.0034}{{\ttfamily 1106.0034}}].

\bibitem{Choudhury:1996ia}
D.~Choudhury and P.~Roy, \emph{{New constraints on lepton nonconserving R-parity violating couplings}}, \href{https://doi.org/10.1016/0370-2693(96)00444-3}{\emph{Phys. Lett. B} {\bfseries 378} (1996) 153} [\href{https://arxiv.org/abs/hep-ph/9603363}{{\ttfamily hep-ph/9603363}}].

\bibitem{Chemtob:2004xr}
M.~Chemtob, \emph{{Phenomenological constraints on broken R parity symmetry in supersymmetry models}}, \href{https://doi.org/10.1016/j.ppnp.2004.06.001}{\emph{Prog. Part. Nucl. Phys.} {\bfseries 54} (2005) 71} [\href{https://arxiv.org/abs/hep-ph/0406029}{{\ttfamily hep-ph/0406029}}].

\bibitem{Cai:2024fkl}
X.~Cai, J.~Li, R.~Ding, M.~Lu, Z.~You and Q.~Li, \emph{{Search for R-Parity-Violation-Induced Charged Lepton Flavor Violation at Future Lepton Colliders}}, \href{https://doi.org/10.3390/universe10060243}{\emph{Universe} {\bfseries 10} (2024) 243} [\href{https://arxiv.org/abs/2403.14993}{{\ttfamily 2403.14993}}].

\bibitem{Langacker:2008yv}
P.~Langacker, \emph{{The Physics of Heavy $Z^\prime$ Gauge Bosons}}, \href{https://doi.org/10.1103/RevModPhys.81.1199}{\emph{Rev. Mod. Phys.} {\bfseries 81} (2009) 1199} [\href{https://arxiv.org/abs/0801.1345}{{\ttfamily 0801.1345}}].

\bibitem{MEG:2016leq}
{\scshape MEG} collaboration, \emph{{Search for the lepton flavour violating decay $\mu ^+ \rightarrow \mathrm {e}^+ \gamma $ with the full dataset of the MEG experiment}}, \href{https://doi.org/10.1140/epjc/s10052-016-4271-x}{\emph{Eur. Phys. J. C} {\bfseries 76} (2016) 434} [\href{https://arxiv.org/abs/1605.05081}{{\ttfamily 1605.05081}}].

\bibitem{SINDRUM:1987nra}
{\scshape SINDRUM} collaboration, \emph{{Search for the Decay $\mu^+ \to e^+ e^+ e^-$}}, \href{https://doi.org/10.1016/0550-3213(88)90462-2}{\emph{Nucl. Phys. B} {\bfseries 299} (1988) 1}.

\bibitem{SINDRUMII:2006dvw}
{\scshape SINDRUM II} collaboration, \emph{{A Search for muon to electron conversion in muonic gold}}, \href{https://doi.org/10.1140/epjc/s2006-02582-x}{\emph{Eur. Phys. J. C} {\bfseries 47} (2006) 337}.

\bibitem{SINDRUMII:1993gxf}
{\scshape SINDRUM II} collaboration, \emph{{Test of lepton flavor conservation in mu ---\ensuremath{>} e conversion on titanium}}, \href{https://doi.org/10.1016/0370-2693(93)91383-X}{\emph{Phys. Lett. B} {\bfseries 317} (1993) 631}.

\bibitem{SINDRUMII:1996fti}
{\scshape SINDRUM II} collaboration, \emph{{Improved limit on the branching ratio of mu ---\ensuremath{>} e conversion on lead}}, \href{https://doi.org/10.1103/PhysRevLett.76.200}{\emph{Phys. Rev. Lett.} {\bfseries 76} (1996) 200}.

\bibitem{SINDRUMII:1998mwd}
{\scshape SINDRUM II} collaboration, \emph{{Improved limit on the branching ratio of mu- --\ensuremath{>} e+ conversion on titanium}}, \href{https://doi.org/10.1016/S0370-2693(97)01423-8}{\emph{Phys. Lett. B} {\bfseries 422} (1998) 334}.

\bibitem{ATLAS:2014vur}
{\scshape ATLAS} collaboration, \emph{{Search for the lepton flavor violating decay Z\textrightarrow{}e\ensuremath{\mu} in pp collisions at $\sqrt{s}$ TeV with the ATLAS detector}}, \href{https://doi.org/10.1103/PhysRevD.90.072010}{\emph{Phys. Rev. D} {\bfseries 90} (2014) 072010} [\href{https://arxiv.org/abs/1408.5774}{{\ttfamily 1408.5774}}].

\bibitem{OPAL:1995grn}
{\scshape OPAL} collaboration, \emph{{A Search for lepton flavor violating Z0 decays}}, \href{https://doi.org/10.1007/BF01553981}{\emph{Z. Phys. C} {\bfseries 67} (1995) 555}.

\bibitem{DELPHI:1996iox}
{\scshape DELPHI} collaboration, \emph{{Search for lepton flavor number violating Z0 decays}}, \href{https://doi.org/10.1007/s002880050313}{\emph{Z. Phys. C} {\bfseries 73} (1997) 243}.

\bibitem{ATLAS:2019pmk}
{\scshape ATLAS} collaboration, \emph{{Searches for lepton-flavour-violating decays of the Higgs boson in $\sqrt{s}=13$ TeV pp collisions with the ATLAS detector}}, \href{https://doi.org/10.1016/j.physletb.2019.135069}{\emph{Phys. Lett. B} {\bfseries 800} (2020) 135069} [\href{https://arxiv.org/abs/1907.06131}{{\ttfamily 1907.06131}}].

\bibitem{CMS:2017con}
{\scshape CMS} collaboration, \emph{{Search for lepton flavour violating decays of the Higgs boson to $\mu\tau$ and e$\tau$ in proton-proton collisions at $\sqrt{s}=$ 13 TeV}}, \href{https://doi.org/10.1007/JHEP06(2018)001}{\emph{JHEP} {\bfseries 06} (2018) 001} [\href{https://arxiv.org/abs/1712.07173}{{\ttfamily 1712.07173}}].

\bibitem{KTeV:2007cvy}
{\scshape KTeV} collaboration, \emph{{Search for lepton flavor violating decays of the neutral kaon}}, \href{https://doi.org/10.1103/PhysRevLett.100.131803}{\emph{Phys. Rev. Lett.} {\bfseries 100} (2008) 131803} [\href{https://arxiv.org/abs/0711.3472}{{\ttfamily 0711.3472}}].

\bibitem{CLEO:2008lxu}
{\scshape CLEO} collaboration, \emph{{Search for Lepton Flavor Violation in Upsilon Decays}}, \href{https://doi.org/10.1103/PhysRevLett.101.201601}{\emph{Phys. Rev. Lett.} {\bfseries 101} (2008) 201601} [\href{https://arxiv.org/abs/0807.2695}{{\ttfamily 0807.2695}}].

\bibitem{BaBar:2021loj}
{\scshape BaBar} collaboration, \emph{{Search for Lepton Flavor Violation in~$\Upsilon (3S)\rightarrow e^{\pm}\mu^{\mp}$}}, \href{https://doi.org/10.1103/PhysRevLett.128.091804}{\emph{Phys. Rev. Lett.} {\bfseries 128} (2022) 091804} [\href{https://arxiv.org/abs/2109.03364}{{\ttfamily 2109.03364}}].

\bibitem{BESIII:2022exh}
{\scshape BESIII} collaboration, \emph{{Search for the lepton flavor violating decay~$J/\psi\to e\mu$}}, \href{https://doi.org/10.1007/s11433-022-1995-0}{\emph{Sci. China Phys. Mech. Astron.} {\bfseries 66} (2023) 221011} [\href{https://arxiv.org/abs/2206.13956}{{\ttfamily 2206.13956}}].

\bibitem{MEGII:2018kmf}
{\scshape MEG II} collaboration, \emph{{The design of the MEG II experiment}}, \href{https://doi.org/10.1140/epjc/s10052-018-5845-6}{\emph{Eur. Phys. J. C} {\bfseries 78} (2018) 380} [\href{https://arxiv.org/abs/1801.04688}{{\ttfamily 1801.04688}}].

\bibitem{Mu3e:2020gyw}
{\scshape Mu3e} collaboration, \emph{{Technical design of the phase I Mu3e experiment}}, \href{https://doi.org/10.1016/j.nima.2021.165679}{\emph{Nucl. Instrum. Meth. A} {\bfseries 1014} (2021) 165679} [\href{https://arxiv.org/abs/2009.11690}{{\ttfamily 2009.11690}}].

\bibitem{COMET:2018auw}
{\scshape COMET} collaboration, \emph{{COMET Phase-I Technical Design Report}}, \href{https://doi.org/10.1093/ptep/ptz125}{\emph{PTEP} {\bfseries 2020} (2020) 033C01} [\href{https://arxiv.org/abs/1812.09018}{{\ttfamily 1812.09018}}].

\bibitem{Mu2e:2014fns}
{\scshape Mu2e} collaboration, \emph{{Mu2e Technical Design Report}},  \href{https://arxiv.org/abs/1501.05241}{{\ttfamily 1501.05241}}.

\bibitem{Langacker:2000ju}
P.~Langacker and M.~Plumacher, \emph{{Flavor changing effects in theories with a heavy $Z^\prime$ boson with family nonuniversal couplings}}, \href{https://doi.org/10.1103/PhysRevD.62.013006}{\emph{Phys. Rev. D} {\bfseries 62} (2000) 013006} [\href{https://arxiv.org/abs/hep-ph/0001204}{{\ttfamily hep-ph/0001204}}].

\bibitem{Buras:2021btx}
A.J.~Buras, A.~Crivellin, F.~Kirk, C.A.~Manzari and M.~Montull, \emph{{Global analysis of leptophilic Z' bosons}}, \href{https://doi.org/10.1007/JHEP06(2021)068}{\emph{JHEP} {\bfseries 06} (2021) 068} [\href{https://arxiv.org/abs/2104.07680}{{\ttfamily 2104.07680}}].

\bibitem{ATLAS:2018mrn}
{\scshape ATLAS} collaboration, \emph{{Search for lepton-flavor violation in different-flavor, high-mass final states in $pp$ collisions at $\sqrt s=13 $ TeV with the ATLAS detector}}, \href{https://doi.org/10.1103/PhysRevD.98.092008}{\emph{Phys. Rev. D} {\bfseries 98} (2018) 092008} [\href{https://arxiv.org/abs/1807.06573}{{\ttfamily 1807.06573}}].

\bibitem{Li:2023lin}
J.~Li, W.~Wang, X.~Cai, C.~Yang, M.~Lu, Z.~You et~al., \emph{{A Comparative Study of Z$^{\prime}$ mediated Charged Lepton Flavor Violation at future lepton colliders}}, \href{https://doi.org/10.1007/JHEP03(2023)190}{\emph{JHEP} {\bfseries 03} (2023) 190} [\href{https://arxiv.org/abs/2302.02203}{{\ttfamily 2302.02203}}].

\bibitem{Madgraph}
J.~Alwall, R.~Frederix, S.~Frixione, V.~Hirschi, F.~Maltoni, O.~Mattelaer et~al., \emph{{The automated computation of tree-level and next-to-leading order differential cross sections, and their matching to parton shower simulations}}, \href{https://doi.org/10.1007/JHEP07(2014)079}{\emph{JHEP} {\bfseries 07} (2014) 079} [\href{https://arxiv.org/abs/1405.0301}{{\ttfamily 1405.0301}}].

\bibitem{Pythia}
T.~Sj\"ostrand, S.~Ask, J.R.~Christiansen, R.~Corke, N.~Desai, P.~Ilten et~al., \emph{{An introduction to PYTHIA 8.2}}, \href{https://doi.org/10.1016/j.cpc.2015.01.024}{\emph{Comput. Phys. Commun.} {\bfseries 191} (2015) 159} [\href{https://arxiv.org/abs/1410.3012}{{\ttfamily 1410.3012}}].

\bibitem{Delphes}
{\scshape DELPHES 3} collaboration, \emph{{DELPHES 3, A modular framework for fast simulation of a generic collider experiment}}, \href{https://doi.org/10.1007/JHEP02(2014)057}{\emph{JHEP} {\bfseries 02} (2014) 057} [\href{https://arxiv.org/abs/1307.6346}{{\ttfamily 1307.6346}}].

\bibitem{Punzi:2003bu}
G.~Punzi, \emph{{Sensitivity of searches for new signals and its optimization}}, {\emph{eConf} {\bfseries C030908} (2003) MODT002} [\href{https://arxiv.org/abs/physics/0308063}{{\ttfamily physics/0308063}}].

\bibitem{Cowan:2010js}
G.~Cowan, K.~Cranmer, E.~Gross and O.~Vitells, \emph{{Asymptotic formulae for likelihood-based tests of new physics}}, \href{https://doi.org/10.1140/epjc/s10052-011-1554-0}{\emph{Eur. Phys. J. C} {\bfseries 71} (2011) 1554} [\href{https://arxiv.org/abs/1007.1727}{{\ttfamily 1007.1727}}].

\bibitem{Hayasaka:2010np}
K.~Hayasaka et~al., \emph{{Search for Lepton Flavor Violating Tau Decays into Three Leptons with 719 Million Produced Tau+Tau- Pairs}}, \href{https://doi.org/10.1016/j.physletb.2010.03.037}{\emph{Phys. Lett. B} {\bfseries 687} (2010) 139} [\href{https://arxiv.org/abs/1001.3221}{{\ttfamily 1001.3221}}].

\bibitem{Banerjee:2022vdd}
S.~Banerjee, \emph{{Searches for Lepton Flavor Violation in Tau Decays at Belle II}}, \href{https://doi.org/10.3390/universe8090480}{\emph{Universe} {\bfseries 8} (2022) 480} [\href{https://arxiv.org/abs/2209.11639}{{\ttfamily 2209.11639}}].

\end{thebibliography}\endgroup






\end{document}